\documentclass[preprint,1p]{elsarticle}
\usepackage{graphicx}
\usepackage{amsmath}
\usepackage{natbib}
\usepackage[usenames]{color}
\usepackage{soul}

\usepackage[normalem]{ulem}

\usepackage[utf8]{inputenc}

\usepackage{amsfonts}
\usepackage{dcolumn}        
\usepackage{amssymb}
\usepackage{bm,amssymb}

\usepackage{bm,fancybox}                	     

\begin{document}

\newcommand{\bea}{\begin{eqnarray}}
\newcommand{\eea}{  \end{eqnarray}}
\newcommand{\bit}{\begin{itemize}}
\newcommand{\eit}{  \end{itemize}}

\newcommand{\be}{\begin{equation}}
\newcommand{\ee}{\end{equation}}
\newcommand{\ra}{\rangle}
\newcommand{\la}{\langle}
\newcommand{\U}{\widetilde{U}}

\def\bra#1{{\langle#1|}}
\def\ket#1{{|#1\rangle}}
\def\bracket#1#2{{\langle#1|#2\rangle}}
\def\inner#1#2{{\langle#1|#2\rangle}}
\def\expect#1{{\langle#1\rangle}}
\def\e{{\rm e}}
\def\proj{{\hat{\cal P}}}
\def\tr{{\rm Tr}}
\def\H{{\hat H}}
\def\Hdag{{\hat H}^\dagger}
\def\Lop{{\cal L}}
\def\Ehat{{\hat E}}
\def\Edag{{\hat E}^\dagger}
\def\Shat{\hat{S}}
\def\Sdag{{\hat S}^\dagger}
\def\Ahat{{\hat A}}
\def\Adag{{\hat A}^\dagger}
\def\U{{\hat U}}
\def\Udag{{\hat U}^\dagger}
\def\Zhat{{\hat Z}}
\def\Phat{{\hat P}}
\def\Op{{\hat O}}
\def\id{{\hat I}}
\def\x{{\hat x}}
\def\P{{\hat P}}
\def\Px{\proj_x}
\def\Pr{\proj_{R}}
\def\Pl{\proj_{L}}


\title{Optimal multicore quantum computing with few interconnects}
\author[add1]{J. Montes}
 \ead{jmontes.3@alumni.unav.es}

\author[add2]{F. Borondo}
 \ead{f.borondo@uam.es}

\author[add3]{Gabriel G. Carlo}
\ead{gabrielcarlo@cnea.gob.ar}
 
 \address[add1]{Department of Statistics, Universidad Carlos III de Madrid, Spain}
 
 \address[add2]{Departamento de Qu\'imica, 
 Universidad Aut\'onoma de Madrid,
 Cantoblanco, 28049--Madrid, Spain}

 \address[add3]{Comisi\'on Nacional de Energ\'ia At\'omica, CONICET, Departamento de F\'isica, 
 Av.~del Libertador 8250, 1429 Buenos Aires, Argentina}

\date{\today}

\begin{abstract}
Noisy intermediate-scale quantum processors have produced a quantum computation revolution in recent times. However, to make further advances new strategies to overcome the error rate growth are needed. One possible way out is dividing these devices into many cores. On the other hand, the majorization criterion efficiently classifies quantum circuits in terms of their complexity, which can be directly related to their ability of performing non classically simulatable computations. In this paper, we use this criterion to study the complexity behavior of a paradigmatic universal family of random circuits distributed into several cores with different architectures. We find that the optimal complexity is reached with few interconnects, this giving further hope to actual implementations in nowadays available devices. A universal behavior is found irrespective of the architecture and (approximately) of the core size. We also analyze the complexity properties when scaling processors up by means of adding cores of the same size. We provide a conjecture to explain the results.
\end{abstract}
\maketitle

\section{Introduction}
 \label{sec:intro}

Recent years have witnessed an extremely fruitful development of quantum 
computation in the form of noisy intermediate-scale quantum (NISQ) processors~\cite{Preskill2018}. Since the first claims of quantum supremacy~\cite{Arute}, just a few years ago until very recent advancements in the use of different technologies~\cite{Madsen} and in solving specific problems~\cite{King}, we experienced a steep scientific progress. But despite very recent breakthroughs in error correcting techniques~\cite{error} we have arrived at a point where the devices growth in size leads to strong difficulties in keeping the fidelities high and error rates sufficiently low. Without solving this problem the quest for practical implementations of quantum advantage becomes challenging. A natural answer is the strategy to divide the quantum devices into smaller cores and connect them in suitable arrays, much as in the classical computers fashion~\cite{multicore,Hetenyi2024}. 
But of course in quantum computing preserving quantum complexity is crucial in 
order to have a useful multicore device. At present, generating entangled states 
between different cores is much more demanding than doing so inside of each core~\cite{interconnects,interconnects2}. 
In this way, there is a communication bottleneck that has led to the development 
of different distribution strategies based mainly on minimizing depth and use of quantum communication~\cite{dist1,dist2,dist3}. Now the question arises of which are the minimum 
requirements for a given architecture to retain the complexity~\cite{work1} needed to perform quantum advantageous computations? And moreover, which is the minimum constraint that 
we have to consider in terms of communication effort at the time of hardware design?

There are several benchmarking methods in order to measure the performance of quantum devices. The  recently introduced majorization criterion~\cite{work1}, allows to evaluate the complexity of random quantum circuits, constructed from different families of quantum gates. By considering the fluctuations of the Lorenz curves calculated with the output probability distributions a complexity indicator is defined that distinguishes between universal and non-universal classes of random quantum circuits at a relatively low cost. This has proven very useful in reservoir quantum computing \cite{qrc1, qrc2} and in the characterization of the complexity of various currently available universal (gate-model) quantum processing units (QPUs) \cite{work2}. In this work we extend its applicability to multicore processors. 

We simulate random quantum circuits from a universal set of gates, which were divided into cores and interconnected via swap operations. We consider several prototypical architectures and core sizes. We have found a universal behavior of complexity measured by means of the majorization criterion. As a matter of fact, an optimal complexity is reached at a relatively small number of interconnects independently of the architecture and to some extent of the core sizes. Also, scaling up a processor by adding cores of the same size proves efficient in this sense.

The article is divided into Sec.~\ref{sec:ModelMethod}, where the details of  the circuits architecture and the complexity criterion used are explained, Sec.~\ref{sec:Results}, in which we analyze the results,  and finally Sec.~\ref{sec:Conclusions}, devoted to the conclusions.

\section{Circuit model and benchmarking method}
\label{sec:ModelMethod}

\begin{figure}[ht]
\centering
\includegraphics[width=0.9\textwidth]{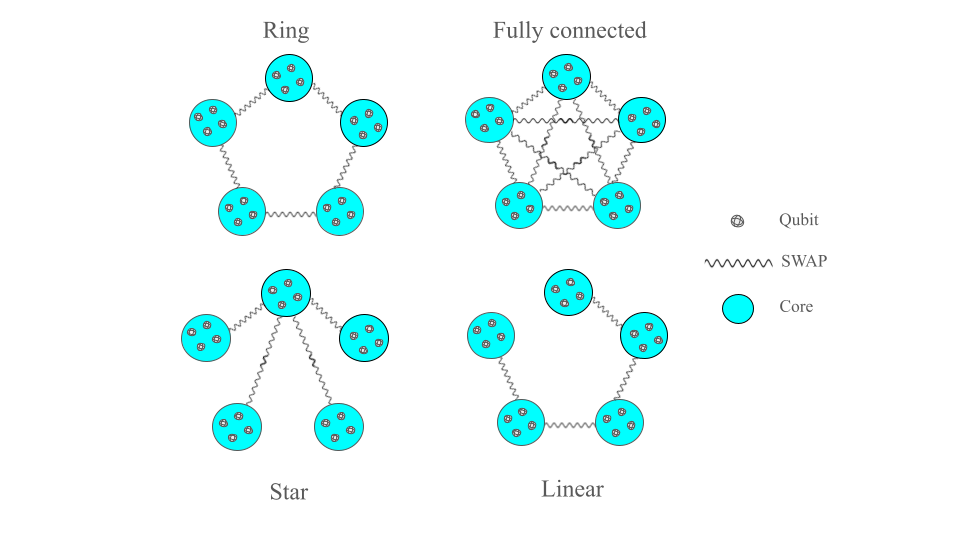}
\includegraphics[width=0.5\textwidth]{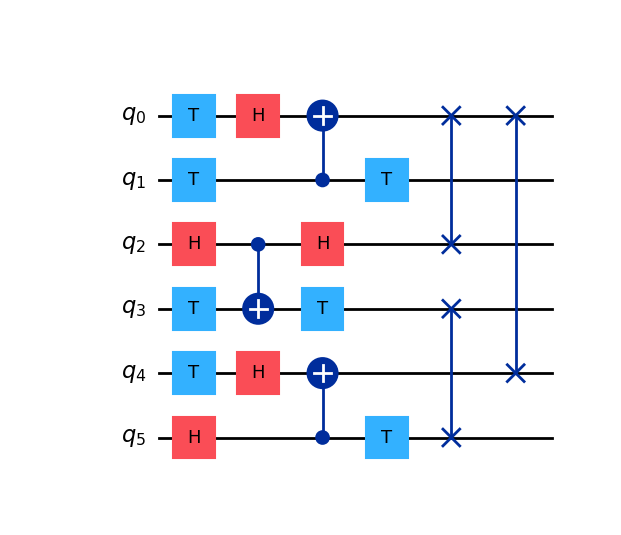}
\hfill
\caption{Upper panel: The different architectures considered in this work. Lower panel: a typical quantum circuit on a ring processor.}
\label{fig0}
\end{figure}

In order to study the complexity behavior of different architectures for distributed quantum computation we first streamline the circuit model. We have considered quantum circuits divided into $N$ cores with four different architectures corresponding to linear, ring, star, and fully connected arrangements. We also took into account the monolithic case, i.e. a full size single core quantum circuit for comparison purposes. The interconnects have been reduced to its barebone version, i.e. swap gates acting on random qubit pairs from different cores, in order to just consider their presence, not their particular features. This allows to single out the pure influence of partitioning on the complexity behavior. We considered different widths of the cores in order to evaluate how the size of individual components affects the overall capacity of the circuit to sustain quantum advantage. We always start  with an initial state $\ket{\psi} = \ket{0} \otimes \ket{0} \otimes ... \otimes \ket{0}$ of $n$ qubits, to which we apply quantum gates from the $G3 = CNOT, H, T$ set (namely the control-NOT, Hadamard and $\pi/8$ phase gates) at random, both at the gate and qubit level (this is a paradigmatic set of quantum gates that is suitable for universal computation in NISQ devices). In Fig.~\ref{fig0} we illustrate the different architectures and a typical quantum circuit for the ring case.

Our benchmarking method, based in the majorization principle, has 
been recently proposed in ~\cite{work1} and we proceed to briefly explain it. Majorization entails ordering vectors according to their components. If two arbitrary vectors $ \textbf{p}, \textbf{q} \in \mathbb{R}^M $ satisfy 
\begin{align}
    \sum_{i=1}^{k} p_{i}^{\downarrow} & \leq \sum_{i=1}^{k} q_{i}^{\downarrow}, \quad 1\leq k < M, \label{eq:major_ineq}\\
    \sum_{i=1}^{M} p_{i} & = \sum_{i=1}^{M} q_{i}\label{eq:normalization},
\end{align}
we say that $\textbf{p} \prec \textbf{q}$, i.e. \textbf{p} is majorized by \textbf{q}, where symbol $^{\downarrow}$ corresponds to sorting the components in non-increasing order. This is a uniformity indicator ~\cite{work2} that we apply to the probabilities of the normalized state vectors after operating on them with a quantum circuit. In Eq.~(\ref{eq:major_ineq}) we compare the $k$-th cumulants of $\textbf{p}$ and $\textbf{q}$ which we denote by $\mathcal{F}_p(k)$ and $\mathcal{F}_q(k)$, respectively. Curves $\mathcal{F}_p(k)$ and $\mathcal{F}_q(k)$ vs $k/N$ are \emph{Lorenz curves} and if \textbf{q} majorizes \textbf{p} the Lorenz curve for \textbf{q} is above the curve for \textbf{p}.

We use the fluctuations of the Lorenz curves corresponding to an ensemble of $n$-qubit random quantum circuits $\{U\}$ to measure its complexity by making them act on an initial state given by $|0\ldots 0\rangle= |0\rangle^{\otimes n}$ and then measuring in the computational basis. The probability distributions thus obtained, $p_U (i) = \left|\bra{0\ldots 0} U \ket{i}\right|^2$ allow us to calculate the cumulants  
$\mathcal{F}_{p_U}(k)$ -- with $k\in\{1,\ldots,2^n\}$ and their fluctuations defined as 
\begin{align}\label{eq:stdF}
    {\rm std}\,[\mathcal{F}_{p_U} (k)] = \sqrt{\langle \mathcal{F}_{p_U}^2 (k) \rangle - \langle \mathcal{F}_{p_U}(k) \rangle^2}.
\end{align}
In this way, our quantum complexity measure is given by the distance of these fluctuations to the ones of $n$-qubit 
Haar-random pure states (denoted ${\rm std}\,[\mathcal{F}_H (k)]$), 
\begin{equation}\label{eq:distancetohaar}
    D_H=\sqrt{\sum_{k=1}^{2^n} \{ {\rm std}\,[\mathcal{F}_{p_U} (k)] - {\rm std}\,[\mathcal{F}_H (k)] \}^2 }.
\end{equation}
The Haar-$n$ curve is a lower limit for universal gate sets \cite{work3} and provides a reference for quantum complexity not attainable by classical means in the large $n$ limit.
    
\section{Results}
 \label{sec:Results}

The main strategy to study the rate at which the maximum complexity associated to 
the corresponding Haar case is reached consists of assigning a fixed number of gates per core ($GPC$) of the multicore quantum processor. After these gates are applied at the local level we use swap gates to interconnect all the cores. We apply just one swap gate between each pair of cores determined by the architecture, with random qubits involved in this exchange. As a consequence, the total number of swaps depends not only on the architecture, but also on the number of gates applied at the core level, as we fix the total number of gates including the swaps ($G$). We take this to be $G=2000$, a safe limit guaranteeing the arrival at the regime of Haar fluctuations. In all cases we initially obtain results for around $200$ gates and then at steps of an approximate size of $100$ gates (the exact numbers will depend on the architecture and $GPC$ choices). 

\begin{figure}[ht]
\centering
\includegraphics[width=1.\textwidth]{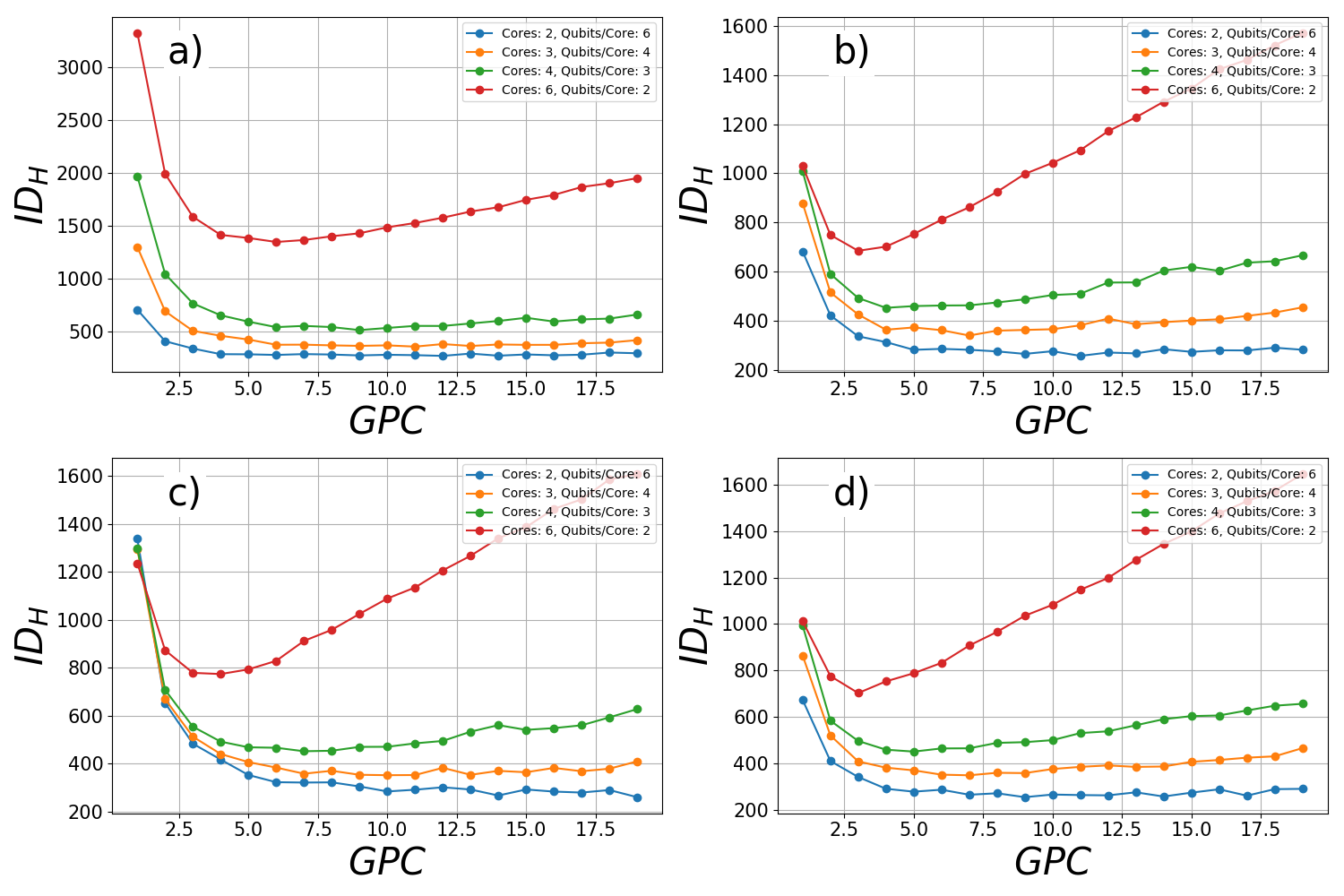}
\hfill
\caption{Integrated Distance to Haar fluctuations ($ID_H$) as a function of the number of gates per core before communications ($GPC$), for the four architectures considered: a) fully connected, b) star, c) ring and d) lineal. We consider 4 different partitions of a 12 qubits circuit (see insets with legends). }
\label{fig1}
\end{figure}

We have calculated the Integrated $D_H$ ($ID_H$) as the integral of the interpolated $D_H$ curves over the total number of gates applied $G$; results as a function of the number of $GPC$ corresponding to the four architectures introduced in Sec.~\ref{sec:ModelMethod} (see upper panel of Fig.~\ref{fig0}) are shown in Fig.~\ref{fig1}. There we have considered a total number of $n=12$ qubits divided into 4 combinations of the  $(N,n_q)$ pair (i.e. (number of cores, qubits per core)), namely $(6,2)$, $(4,3)$, $(3,4)$ and $(2,6)$ (see caption for further details).   
The first thing we notice is that for all the architectures, there is a minimum corresponding to the optimal compromise between local complexity generation at the core level and its distribution via interconnections. Too few local gates fail to generate enough local complexity while too much of them reduce interconnections in excess. Next, the more divided into smaller cores the circuit is, the less efficient in terms of reaching optimal complexity it becomes (greater $ID_H$). This is an expected result since local complexity is less significant at global level and proportionally more resources are devoted to distributing it. The latter argument also explains why the fully connected case separates from the other architectures in general and more markedly for the smaller cores case. We also see that the optimal number of $GPC$ are well defined for the 2 qubits cores case, i.e. the minima of the curves are sharp. These minima become flatter as the size of the cores increase, this being independent of the architecture and strongly related to the local generation of complexity. As the size of the core grows, more $GPC$ are needed but once the minimum $ID_H$ is reached it remains approximately so for a wide range. This allows us to conclude that reaching optimal complexity at a high efficient rate is feasible with a relatively small number of interconnects. 

\begin{figure}[ht]
\centering
\includegraphics[width=1.\textwidth]{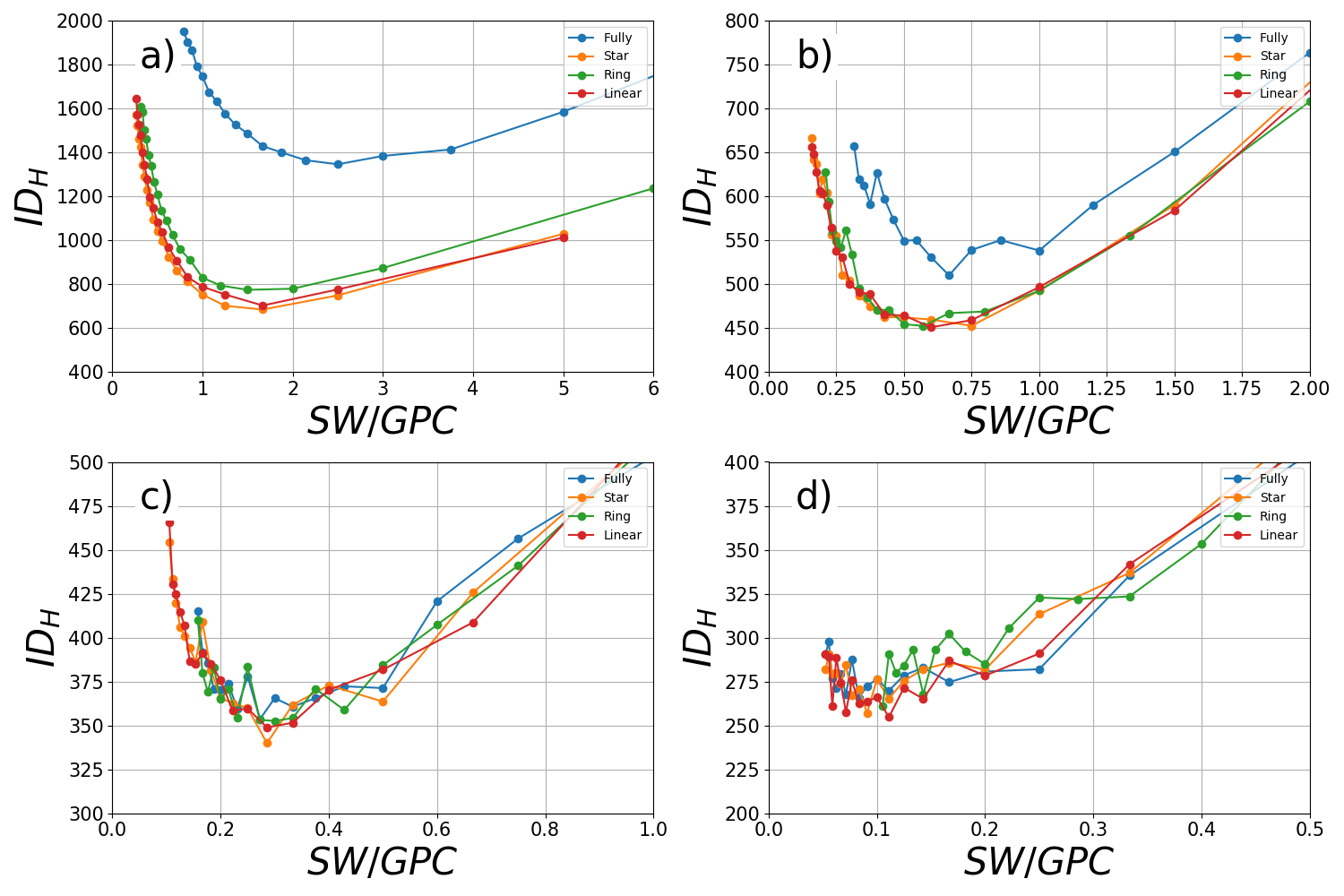}
\hfill
\caption{$ID_H$ as a function of $SW/GPC$ (ratio of swaps over $GPC$), for the four architectures considered (see insets with legends). We consider 4 different partitions of a 12 qubits circuit: a) $(6,2)$, b) $(4,3)$, c) $(3,4)$, and d) $(2,6)$.}
\label{fig2}
\end{figure}

In order to clarify this finding, in Fig.~\ref{fig2} we show the $ID_H$ as a function of the ratio of swaps over $GPC$ (before each set of interconnections), namely $SW/GPC$ and also in a different 
arrangement, grouping together the curves corresponding to the same $(N,n_q)$ for all architectures. 
Examining Fig.~\ref{fig2} we notice that with the exception of the fully connected architecture in the smaller core size, all curves scale to the same behavior in terms of $SW/GPC$. This amounts to saying that for medium to large core sizes compared to the whole processor the way of reaching optimal complexity is essentially architecture independent (keeping in mind the limitation imposed by the simulatable system sizes: for just $2$ cores all architectures naturally coincide and for $3$ star and linear are also the same). Moreover, once a critical swap number relative to the number of gates inside each core is attained, the efficiency does not change significantly. For example, for the $(4,3)$ case once we have $SW/GPC \simeq 0.4$, the $ID_H$ remains well below the $500$ level, up to $SW/GPC \simeq 0.8$. Beyond that point the lack of local complexity generation starts to dominate, but the important result is that it is useless to duplicate the number of swaps with respect to the minimum at which $ID_H \simeq 400$. For the $(3,4)$ case this is even more evident, where $ID_H < 375$ from $SW/GPC \simeq 0.2$ and keeps that way (in a narrow band) until around more than double this value ($0.5$). For the case of $2$ cores (same architecture), $ID_H$ values remain below $300$ from $SW/GPC \simeq 0.05$ to $SW/GPC \simeq 0.2$. In general, it is expected that less swaps are needed as the number of cores in the processor decreases, but the universal scaling obtained and the relative insensitivity to the number of cores in a wide range of $SW/GPC$ are two remarkable results. 

\begin{figure}[ht]
\centering
\includegraphics[width=1.\textwidth]{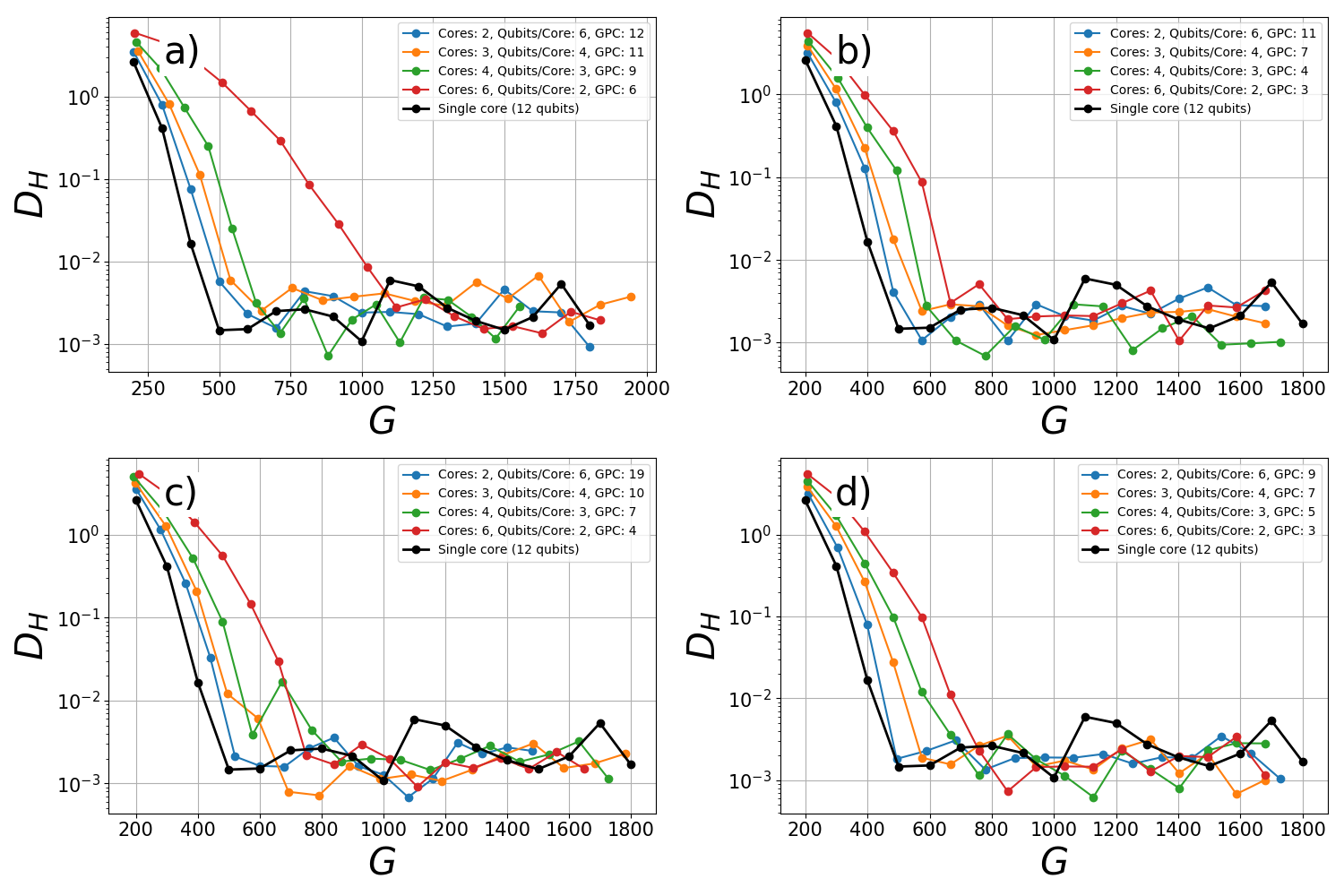}
\hfill
\caption{Distance to Haar fluctuations ($D_H$) as a function of the total number of gates ($G$) applied for the minima seen in Fig.~\ref{fig1} (we use the same panel assignment). Black curves correspond to a single core processor of the same size.}
\label{fig3}
\end{figure}

To conclude this analysis, we show in Fig.~\ref{fig3} the $D_H$ curves as a function of the total number of gates $G$, corresponding to the minima of $ID_H$ found in Fig.~\ref{fig1}. 
Here, we see the way in which the distance to the corresponding Haar fluctuations saturates (we confirm that this saturation is actually reached), and also compare to the monolithic processor case. In fact, with the only exception of the fully connected architecture for the smaller core size of just $2$ qubits, the evolution corresponding to the minima of Fig.~\ref{fig1} shows the same behavior as the single core case. There are no qualitative differences due to partitioning the processor in any architecture whatsoever. This leads us to conjecture an equivalent behavior in terms of the available Hilbert space covering irrespective of these features.

\begin{figure}[ht]
\centering
\includegraphics[width=1.\textwidth]{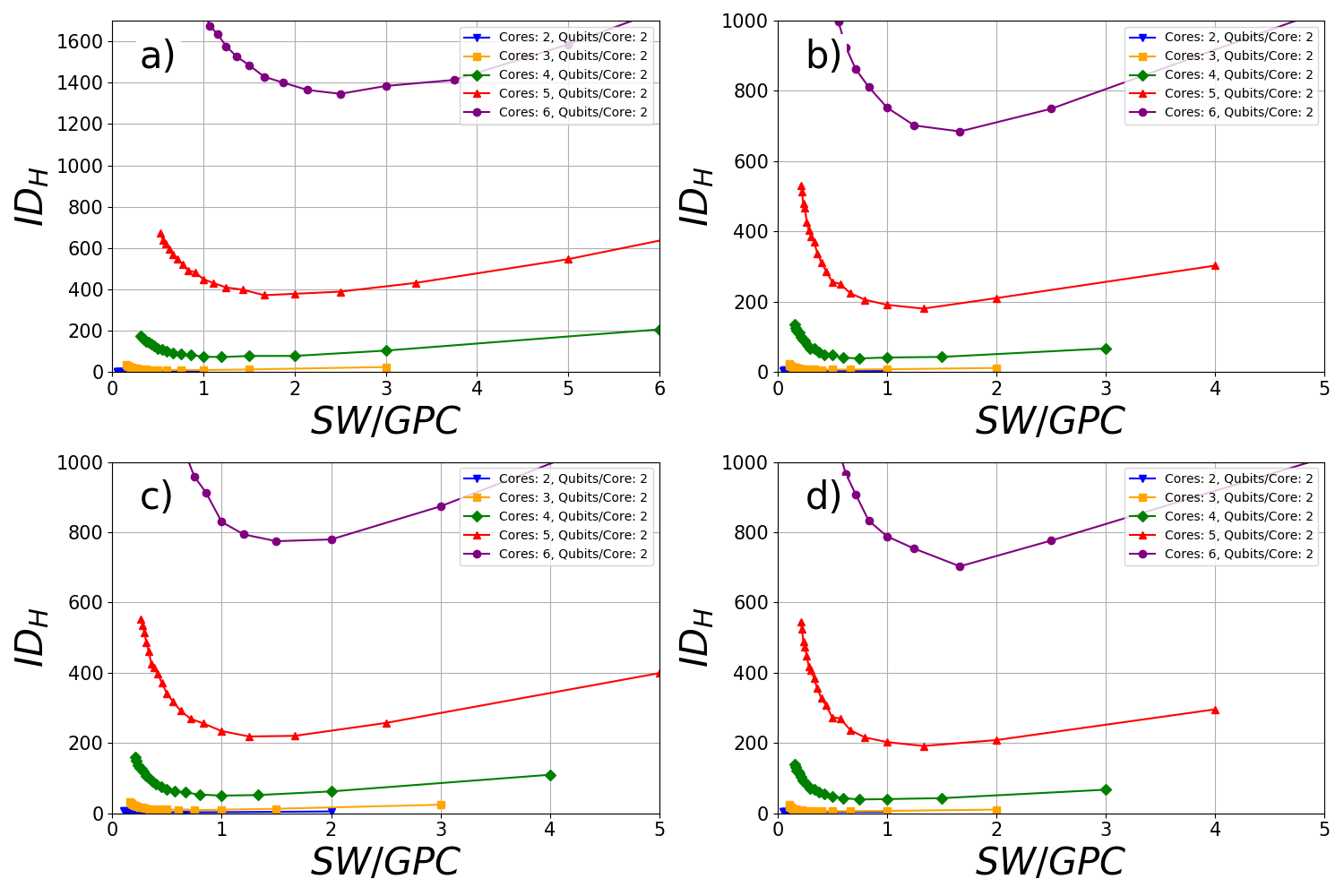}
\hfill
\caption{$ID_H$ as a function of $SW/GPC$, for the four architectures considered (same panel assignment as in Fig.~\ref{fig1}). We consider a growing number of 2 qubits cores in each case (see insets with legends).}
\label{fig4}
\end{figure}

\begin{figure}[ht]
\centering
\includegraphics[width=1.\textwidth]{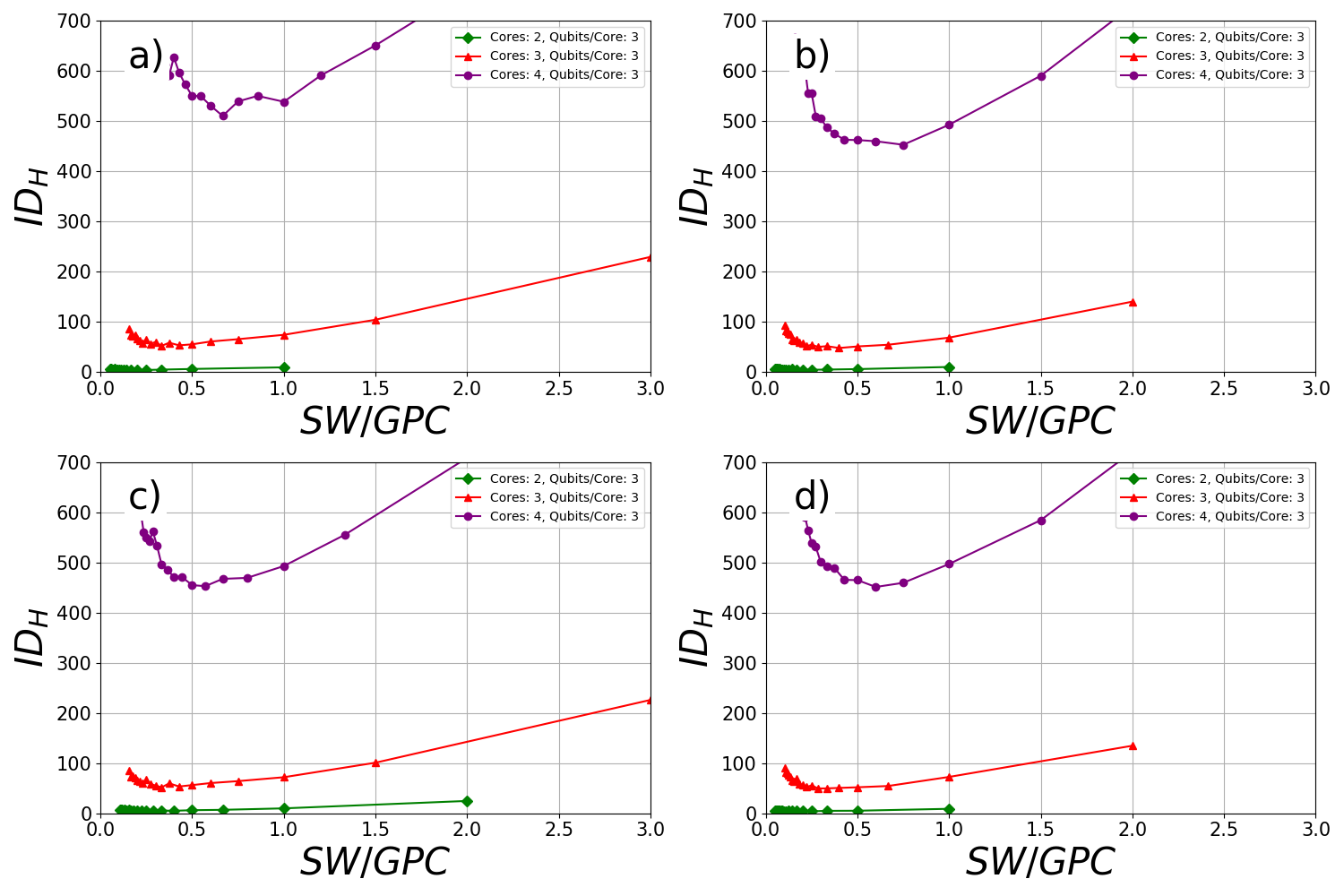}
\hfill
\caption{$ID_H$ as a function of $SW/GPC$, for the four architectures considered (same panel assignment as in Fig.~\ref{fig1}). We consider a growing number of 3 qubits cores in each case (see insets with legends).}
\label{fig5}
\end{figure}

\begin{figure}[ht]
\centering
\includegraphics[width=1.\textwidth]{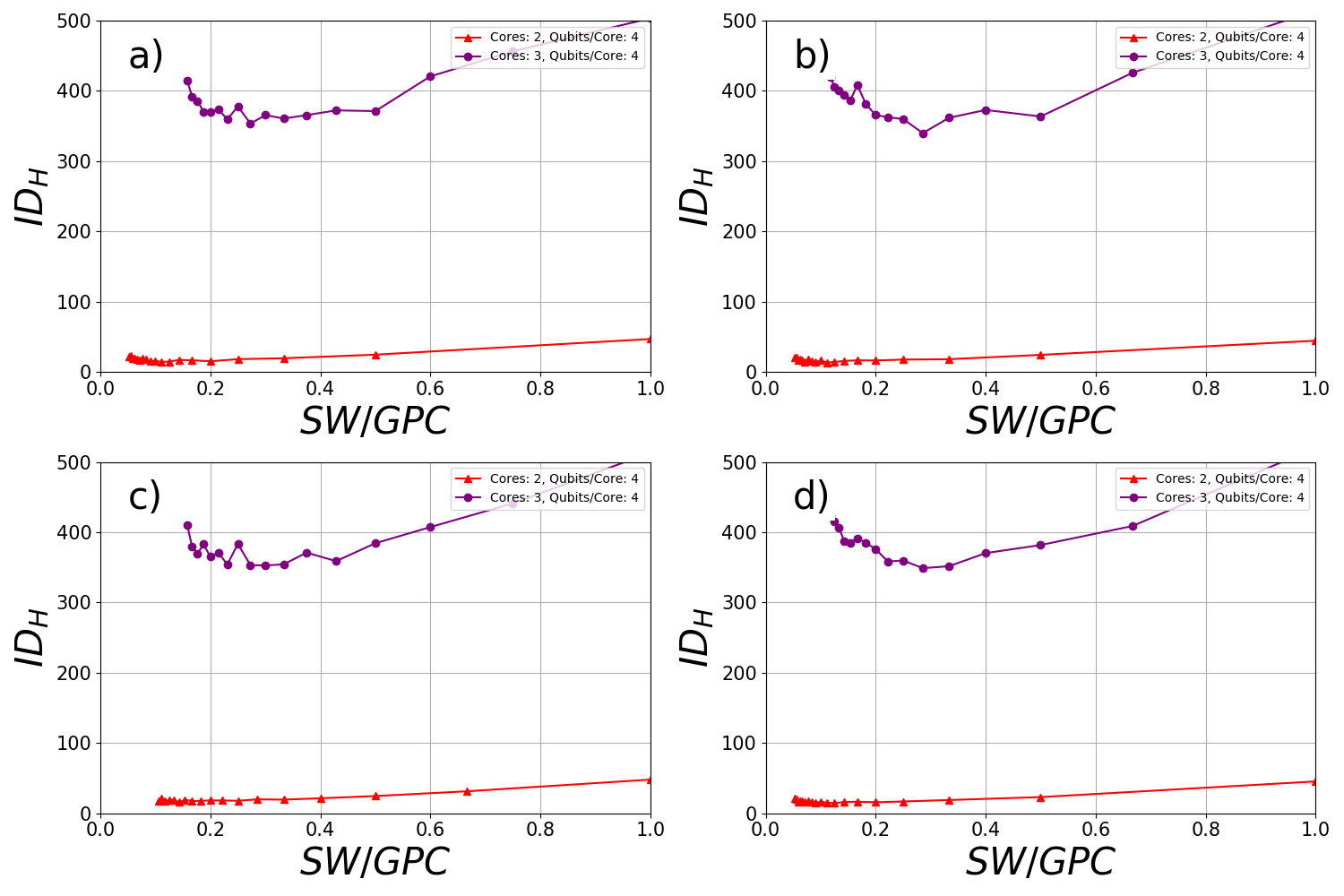}
\hfill
\caption{$ID_H$ as a function of $SW/GPC$, for the four architectures considered (same panel assignment as in Fig.~\ref{fig1}). We consider a growing number of 4 qubits cores in each case (see insets with legends).}
\label{fig6}
\end{figure}

To complete the picture we have extended our analysis in order to characterize the 
complexity behavior of a processor scaled up by means of adding cores of the same size. Results can be seen in Figs. \ref{fig4}, \ref{fig5}, and \ref{fig6}, which correspond 
to the cases of cores of $2$, $3$ and $4$ qubits, respectively. 
The first thing we notice is that the minima of $ID_H$ become less flat as the number of 
cores grow for all core sizes and architectures, this being more evident for smaller cores.  The fully connected architecture reaches higher minima in all cases with more than $3$ cores, confirming the behavior previously noticed. In the $2$ qubits case the range of $SW/GPC$ for which optimal complexity is reached stretches for one order of magnitude for the smaller processors ($2$ and $3$ cores) and shrinks to around a factor of $2$ for larger ones. 
In the $3$ qubits case this is also found, while for the $4$ qubits scenario the minima is flat over half an order of magnitude for the $2$ and $3$ cores considered. This makes us think that scaling greater cores will keep its efficiency with a relatively lower number of interconnects, but greater systems need to be simulated in order to draw further conclusions about this scaling.
 
Here we present a conjecture in order to explain our main result, i.e. the need of relatively few interconnects to reach optimal complexity in multicore processors. The idea is to relate the complexity generation 
to a map acting over a tensor $\gamma$ of rank equal to the number of cores $N$. We define its elements by \(\gamma_{i_1,i_2,\dots,i_{N}}=\Psi_{i_1 i_2 \dots i_{N}}\) where the $\Psi$'s are the coefficients of the state vector in the computational basis expanded by $|i_1 i_2 \dots i_{N}\rangle$. For example, the matrix corresponding to the simplest case of $(N,n_q)=(2,2)$ reads
\[
\gamma = \begin{pmatrix}
\Psi_{0000} & \Psi_{0001} & \Psi_{0010} & \Psi_{0011} \\
\Psi_{0100} & \Psi_{0101} & \Psi_{0110} & \Psi_{0111} \\
\Psi_{1000} & \Psi_{1001} & \Psi_{1010} & \Psi_{1011} \\
\Psi_{1100} & \Psi_{1101} & \Psi_{1110} & \Psi_{1111}
\end{pmatrix}.
\]
The action of the considered quantum gates on this matrix can be grouped into 3 kinds 
\begin{enumerate}
    \item \textbf{Exchange}: these are the CNOT and SWAP gates which  exchange certain elements of the matrix.
    \item \textbf{Phase change}: these are the T gates that add a phase \(e^{i\phi}\) to certain elements.
    \item \textbf{Mix}: these are the H gates, which mix certain elements.
\end{enumerate}

\begin{figure}[h]
    \centering
    \includegraphics[width=1.1\textwidth]{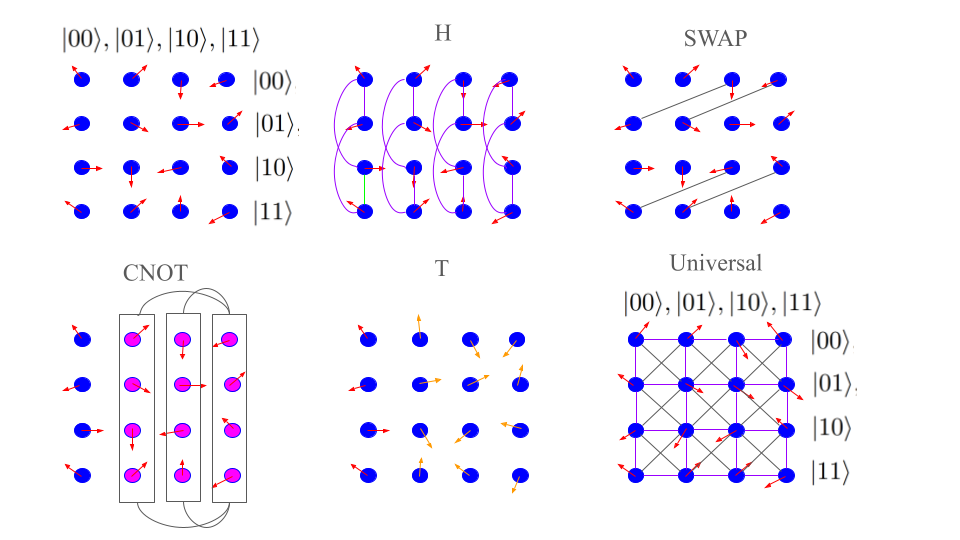}
    \caption{Effect of quantum gates on the matrix elements: (a) Original matrix; (b) Matrix after applying the \( H \) gate; (c) Effect of the SWAP gate connecting diagonal elements; (d) Effect of the \( \text{CNOT} \) gate permuting columns; (e) Effect of the \( T \) gate changing elements phases; (f) The combined action of all gates.}
    \label{fig:nodes}
\end{figure}

All this is illustrated in Fig.~\ref{fig:nodes}
The effect of applying the \( T \) gate is to change the phase of two rows or columns, depending on which core it belongs to. Since \( T \) is applied to a particular qubit, we can have the options \( T_1, T_2, \ldots, T_4 \), affecting columns 2 and 4 or 3 and 4. In the figure, the phase of three columns has been changed to illustrate all the possibilities of the \( T \) gate acting on the qubits of a given core.  The effect of the \( H \) gate is to mix rows or columns: 1-2, 3-4, or 1-3, 2-4. Then, the \( \text{CNOT} \) gate permutes entire columns or rows, again depending on which core is applied to, such as 2-4 or 3-4. Finally the SWAP operations exchange some elements diagonally.

\begin{figure}[h]
    \centering
    \includegraphics[width=1.1\textwidth]{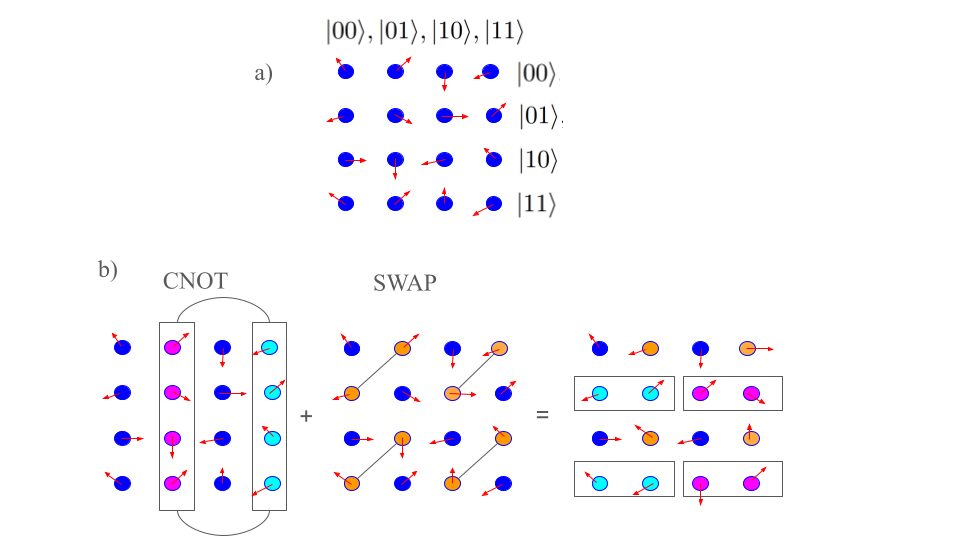}
    \caption{(a) Initial matrix. (b) The combined effect of applying a \textit{CNOT} gate within a single core followed by a \textit{SWAP} operation between two cores, effectively resulting in a row column exchange.}
    \label{fig:suma}
\end{figure}

The combination of \textit{CNOT} and \textit{SWAP} gates is a key ingredient in this mapping of the matrix in which we have arranged 
our computational state. It distributes entanglement along the whole qubit set, and this is represented as an exchange of rows and columns as seen on Fig.~\ref{fig:suma}.

This action can be interpreted as the mixing of a chaotic map. The SWAP gates are responsible for a sort of stretching and folding that resembles that of the baker map, a streamlined model of the Smale's horseshoe \cite{Arnold}. Local generation of complexity amounts to mixing and rearranging rows and columns separately. This involves many matrix elements at once and quickly saturates (at an order $2^{n_q}$ operations). Then, with a relatively small amount of stretching and folding, this map spreads local complexity over the whole matrix.

\section{Conclusions}
 \label{sec:Conclusions}

By analyzing different ways of partitioning a fixed number of qubits into several cores giving rise to a multicore quantum processor we were able to identify a universal behavior irrespective of the architecture and the size of the cores for intermediate cases (i.e. cores neither comprised of too few or too many qubits compared to the total available number of them). In fact, the optimal compromise between the local complexity generation at the core level and the distribution over the whole processor by means of simple swap operations is governed by just the $SW/GPC$ ratio. On the other hand the optimum of the $ID_H$ is reached for relatively small values of $SW/GPC$ and stays approximately so for a few multiples of it. These two findings lead us to the conclusion that optimal complexity is always attainable with relatively few interconnects.

We have also studied how the scaling of the quantum processor by means of adding more cores of the same size affects the complexity behavior. 
We have found that for smaller cores the optimal complexity is reached less efficiently and at the expense of a greater number of interconnects. We found hints that greater cores can be scaled up with a relatively lower number of connections among them, but more calculations are needed to be conclusive about this point.

In future work we will elaborate our conjecture about the efficient action of the SWAP gates. On the other hand, we will consider the evaluation of more realistic interconnects given by teleportation of qubits. Also, we plan to study the effect of noise and implement other complexity measures and circuit kinds.

\section{Acknowledgments}
This work has been partially supported by the Spanish Ministry of Science, 
Innovation and Universities, Gobierno de Espa\~na, under Contract No.\ PID2021-122711NB-C21. Support from CONICET is greatfully acknowledged.

%
\end{document}